\documentclass[prd,aps]{revtex4}

\newcommand{\bbA}{\mbox{\boldmath $A$}}
\newcommand{\bbB}{\mbox{\boldmath $B$}}

\begin{document}

\title{Exploiting gauge and constraint freedom in 
hyperbolic formulations of Einstein's equations}
\author{Olivier Sarbach}
\email{sarbach@phys.lsu.edu}
\author{Manuel Tiglio}
\email{tiglio@lsu.edu}
\affiliation{Department of Physics and Astronomy, Louisiana State
University, 202 Nicholson Hall, Baton Rouge, LA 70803-4001}

\begin{abstract}
We present new many-parameter families of strongly and symmetric
hyperbolic formulations of
Einstein's equations that include quite general algebraic and live
 gauge conditions for the lapse. 

The first system that we present has $30$ variables and
incorporates an algebraic relationship between the lapse and
the determinant of the three metric that generalizes the densitized
lapse prescription. The second system has $34$ variables and uses a family of live
gauges that generalizes the Bona-Masso slicing conditions.  

These  systems have free parameters even after imposing 
hyperbolicity and are expected to be useful in 3D numerical
evolutions. We discuss under what conditions there are no 
superluminal characteristic speeds. 

\end{abstract}

\maketitle

\section{Introduction}
The use of hyperbolic formulations of Einstein's equations in
numerical simulations has several advantages (see \cite{rev} for reviews).   
In particular this includes the
ability of giving boundary conditions that are consistent with both
the evolution equations and the constraints \cite{bound}. At the 
discrete level well posedness supplemented by 
consistency ---in order to make sure one is solving Einstein's
equations--- implies that one can find discretizations that are
numerically stable in the sense of the Lax theorem, a fact that is
 otherwise not necessarily true \cite{cal_conv}. 

Well posedness guarantees that there is a bound
in the growth of the solutions that is independent of the initial
data, but it does allow this
bound to grow with time, and even fast. If this happens, numerical solutions
can grow fast as well and, since Einstein's equations are non-linear,
this can make the code crash in a finite time. This is 
what usually happens in 3D black hole
simulations. This growth might appear due to several factors,
including fast growing gauge or constraint violating modes. 

Initially, hyperbolic formulations of Einstein's equations with
associated well posed initial-value problems  
relied on the use of
harmonic coordinates \cite{ch}, that is, 
$\nabla ^{\alpha }\nabla _{\alpha} x^{\mu}=0$. Although these
coordinates have been used recently (see \cite{gar} and
references therein, and also \cite{jeff}), they might be  
too restrictive for numerical applications, since one 
might want to use another gauge which is better suited to the dynamics
of a given numerical evolution. Initial efforts towards
relaxing the harmonic condition 
introduced hyperbolic formulations that 
used  time harmonic coordinates,
i.e. $\nabla ^{\alpha }\nabla _{\alpha} t =0$ (see \cite{rev,fri} and
references therein). This condition
  can be written as
\begin{displaymath}
\partial_t N - \beta^i\partial_i N = - N^2K \;,
\end{displaymath}
where $K$ is the trace of the extrinsic curvature, $\beta ^i$ the
shift vector, and $N$ the lapse.  A generalization of the time
harmonic slicing,  
of the form
\begin{equation}
\partial_t N - \beta^i\partial_i N = - N^2f(N)K
\label{bm}
\end{equation}
with $f>0$ was introduced by Bona and Masso (BM) \cite{bona1} \footnote{Actually,
there are two BM strongly hyperbolic formulations, the first one uses time harmonic
slicing \cite{bona}, and the second one \cite{bona1} uses conditions
(\ref{bm}). Along this paper, when we refer to the BM formulation, it
will be to the second one.}, being able to
incorporate this condition in a
strongly hyperbolic (SH) evolution system. Here by SH we mean that the
principal part has a complete set of eigenvectors with real
eigenvalues. If the principal part can be diagonalized with a
transformation that is uniformly bounded and smooth in all of its
arguments, then the initial value problem can be shown to be well
posed (this will be discussed later). The BM formulation 
 represents the first effort in moving
away from the
time harmonic slice while achieving strong hyperbolicity, incorporating other 
gauge conditions often used in numerical relativity. For example, (\ref{bm})
includes the ``$1+\log$'' condition ($f=a/N$, with $a$ some
constant). In fact, given a hyperbolic  
 formulation of
Einstein's equations that introduces the lapse as a dynamical variable
through condition (\ref{bm}), one can add to the right hand side (RHS) of
this equation any function $S$ of spacetime, 
\begin{equation}
\partial_t N - \beta^i\partial_i N =  - N^2f(N)K + S(x^{\mu}) \; ,
\label{gral_harm}
\end{equation}
without changing the
level of hyperbolicity of the whole system. From this
observation it is clear that any spacetime with given lapse and shift
 satisfies equation (\ref{gral_harm}), provided $S$ is chosen
appropriately (namely, $S= \partial_t
N - \beta^i\partial_i N + N^2f(N)K $). 

A closely related slicing condition, which has also been used in several
 hyperbolic formulations (see \cite{rev,fri,kst} and references
 therein), is obtained by 
densitizing the lapse. That is, one writes 
\begin{equation}
N=e^Qg^{\sigma } \;,
\label{dens1}
\end{equation}
where $Q=Q(x^{\mu})$ is an arbitrary but a priori specified function
of spacetime, $g$ is the determinant
of the three-metric and $\sigma $ any constant (strong hyperbolicity
implies $\sigma >0$). As before, given any
 spacetime with lapse and shift 
there is always a $Q$ that satisfies eq.(\ref{dens1}). Restricting to the case of zero
 shift and taking a time
derivative of both sides of Eq.(\ref{dens1}) one gets
\begin{displaymath}
\partial_t N  =  - 2\sigma N^2K + N\partial_t Q \; , \label{harm2}
\end{displaymath}
which shows that in this case condition (\ref{dens1}) is  time
harmonic if $\sigma =1/2$ and $\partial _t Q=0$.

Most hyperbolic formulations use either a densitized lapse or some of
the BM conditions. The Kidder-Scheel--Teukolsky (KST) formulation \cite{kst}, for
example, uses densitized lapse. This formulation is a many-parameter 
generalization of previous systems
\cite{prev} 
that is SH or symmetric hyperbolic if these free parameters
satisfy certain inequalities. KST have shown that one can make
use of the freedom in these parameters to extend the lifetime of  3D
numerical 
evolutions of black holes. 

As already
discussed, one can always evolve a known spacetime in a given slicing 
by using the appropriate densitization $Q$. Sometimes one can
even use the same function $Q$ to evolve a high, non-linear distortion
of the same spacetime 
\cite{cal1}. But it is not clear what densitization to choose if
one aims to evolve a spacetime that describes, say, the collision
of two black holes. This is a rather general problem of gauge
prescription. For example, even though there are some features of
the BM slicings that are already understood (like its singularity
avoiding properties), 
it is still far from clear what the ``most appropriate choice'' for
the function $f(N)$ is 
(see section \ref{discussion}). However, it seems one could benefit from 
more general slicing conditions in 
the same way the free
parameters of the KST system turn out to be useful even though one does
not completely understand why.

The aim of this paper is to present families of SH and symmetric 
formulations of Einstein's equations that combine both the freedom of
the KST system  with a family of lapse choices that includes those described above.

The first family that we present is a straightforward generalization
of the KST one that has, instead of a densitized lapse, an algebraic
relation of the form 
\begin{equation}
N=N(g,x^{\mu})\; .
\label{alg_lapse}
\end{equation}
That is, the lapse still
depends on the spacetime coordinates and the determinant of the
three-metric in an a
priori prescribed way, but now the power law dependency on $g$ is
relaxed. The motivation for introducing this generalization is that it
is closely related to the BM conditions, in the same way densitized
lapse is related to a time harmonic slicing. More specifically, 
condition (\ref{alg_lapse}) is equivalent to a particular case of condition (\ref{st}) 
 presented below
provided the function $F$ in Eq.(\ref{st}) satisfies $\partial _K F>0$ or,
equivalently, 
$\partial_g N>0$. As we will 
see later, these two conditions are necessary for strong 
hyperbolicity. Therefore, in those cases there will be a
 correspondence between the live gauges and their
algebraic counterpart  
 (though the resulting
evolution equations will not necessarily be 
equivalent off the constraint surface). 
 This family has, as the KST one, $30$ variables and 
is presented in
Section \ref{alg_section}. There we also perform a characteristic analysis in
order to give sufficient conditions for strong hyperbolicity to hold
and in order to analyze under what conditions there are no
superluminal speeds. 

In section 
\ref{live_section} we present a many-parameter family of SH
 formulations that includes the following live gauge conditions
\begin{equation}
\partial_t N - \beta^i\partial_i N = - NF(N,K,x^{\mu})  \; ,
\label{st}
\end{equation}
where $F$ is any function of its arguments that satisfies  $\partial
_K F>0$. These are
generalizations of the BM ones that relax the linear
dependency on $K$. Since now there are evolution equations for the  lapse and
its spatial derivatives, our system has $34$ variables. This system
differs from the BM 
formulation in various aspects. For example, it uses different 
variables (in fact, the BM formulation 
introduces three extra variables in addition to the ones we consider here).

In section \ref{symm} we will explicitly show that some of the
formulations presented in this paper are symmetric
hyperbolic; well posedness of the initial value problem
follows immediately for these cases. These symmetric hyperbolic
systems  are obtained by
restricting some of the free
parameters of our formulations, but not the gauge choice. That is, we are
able to get symmetric hyperbolic systems with lapses given by
Eqs. (\ref{alg_lapse}) or (\ref{st}).

\section{Strongly hyperbolic formulations with an algebraic gauge}
\label{alg_section}

The system presented in this section consists of the $30$ variables 
$\{g_{ij}, K_{ij}, d_{kij}\}$, where $g_{ij}$ is the
three-metric, $K_{ij}$ the extrinsic curvature, and where the
extra variables $d_{kij}$ are equal to the first order spatial
derivatives $\partial_k g_{ij}$ of the three-metric provided
the constraints are satisfied.

The evolution equations in vacuum are
obtained by adding constraints to the RHS of the evolution 
equations obtained from setting to zero the four-dimensional Ricci tensor.
Following the notation of KST
(except that here we define $\partial_0 \equiv (\partial_t - \pounds
_{\beta })/N$),
\begin{eqnarray}
\partial_0 g_{ij} &=&  -2K_{ij} \; ,
\label{gdot}\\
\partial_0 K_{ij} &=& R_{ij} - \frac{1}{N} \nabla_i\nabla_j N - 
2 K_{ia} K^a_{\; j} + K K_{ij}
+ \gamma\, g_{ij}C + \zeta\, g^{ab}C_{a(ij)b} \; ,
\label{kdot}\\
\partial_0 d_{kij} &=& -2\partial_k K_{ij} - 2\frac{\partial_k N}{N} K_{ij}
+ \eta\, g_{k(i}C_{j)} + \chi\, g_{ij}C_k \; ,
\label{ddot}
\end{eqnarray}
where $\{\gamma, \zeta, \eta, \chi \}$ are free parameters, 
$C=(R-K_{ab}K^{ab}+K^2)/2$ is the 
Hamiltonian constraint, $C_i = \nabla ^aK_{ai} -\nabla_i K$ the momentum one, and
$C_{kij} = d_{kij}-\partial_k g_{ij}$, $C_{lkij} = \partial_{[l}d_{k]ij}$
are constraints that arise due to the introduction of the extra
variables. The Ricci tensor $R_{ij}$ belonging to the
three-metric is written as
\begin{displaymath}
R_{ij} = \frac{1}{2} g^{ab} \left(
 -\partial_a d_{bij} + \partial_a d_{(ij)b} + \partial_{(i} d_{|ab|j)} - \partial_{(i} d_{j)ab} \right)
 + \frac{1}{2} d_i^{\;\; ab} d_{jab} + \frac{1}{2}(d_k - 2b_k)\Gamma^k_{\; ij}
 - \Gamma^k_{\; lj} \Gamma^l_{\; ik}\; ,
\end{displaymath}
where $b_j \equiv d_{kij} g^{ki}$, $d_k \equiv d_{kij} g^{ij}$ and
\begin{displaymath}
\Gamma^k_{\; ij} = \frac{1}{2} g^{kl} \left( 2d_{(ij)l} - d_{lij} \right).
\end{displaymath}
Similarly, the momentum constraint gives
\begin{displaymath}
C_i = g^{ab}\left( \partial_a K_{bi} - \partial_i K_{ab} \right) 
    + \frac{1}{2}(d^k - 2b^k)K_{ki} + \frac{1}{2} d_i^{\;\; ab} K_{ab}\; .
\end{displaymath}

The shift is assumed to be a prescribed function of spacetime. 
In contrast to KST, where the lapse is defined by Eq. (\ref{dens1}), here
we will consider the lapse to be 
an arbitrary function of the coordinates and the determinant of the
three metric, 
 $N=N(g,x^{\mu}$) (as described below, strong hyperbolicity requires 
$\partial_g N > 0$). 
In this case, we have
\begin{eqnarray}
\frac{\partial _k N}{N}&=&  \sigma _{eff}d_k + \frac{\partial
N}{N\partial x^k}\; , \nonumber\\
\frac{1}{N}\nabla_i\nabla_j N &=& \sigma_{eff}\left( 
 g^{ab}\partial_{(i} d_{j)ab} - d_i^{\;\; ab} d_{jab} + d_i d_j \right)
  + \frac{1}{N}\left( \frac{\partial ^2 N}{\partial g ^2} g^2 d_i d_j
+ g\frac{\partial^2 N}{\partial
g\partial x^i} d_j
  + g\frac{\partial^2 N}{\partial g\partial x^j} d_i +
\frac{\partial^2 N}{\partial x^i\partial x^j} \right)
\nonumber\\
 &-& \Gamma^k_{\; ij}\left( \sigma_{eff} d_k + \frac{\partial
N}{N\partial x^k} \right),
\nonumber
\end{eqnarray}
where the ``effective'' $\sigma$ is defined by
\begin{displaymath}
\sigma _{eff}:= gN^{-1}\frac{\partial N}{\partial g} 
\end{displaymath}
($\sigma _{eff} $
coincides with $\sigma $ in Eq.(\ref{dens1}) if the lapse is
densitized). In order to analyze hyperbolicity one has to look at the
principal part of the system, that is, the terms that have spatial
derivatives of the main variables. In this case the principal part is
 \begin{eqnarray}
\partial_0 g_{ij}&=&  l.o. \, , 
\label{Eq:gij}\\
\partial _0 K_{ij}&=& \frac{1}{2} g^{ab}\left( 
  - \partial_a d_{bij} + (1+\zeta )\partial_a d_{(ij)b} 
  + (1-\zeta)\partial_{(i} d_{|ab|j)} - \left(1 + 
2\sigma_{eff}\right)\partial_{(i}d_{j)ab} 
  + \gamma\, g_{ij} g^{kl}\partial_a(d_{klb}-d_{bkl}) \right) + l.o. \, , \quad\;\;
\label{Eq:Kij}\\
\partial _0 d_{kij}&=& -2\partial _kK_{ij} 
 + \eta\,g_{k(i}g^{ab}(\partial_{|a|} K_{j)b} - \partial_{j)} K_{ab}) 
 + \chi\,g_{ij}g^{ab}(\partial_a K_{kb} - \partial_k K_{ab}) + l.o. \, , 
\label{Eq:dkij}
\end{eqnarray}
where $l.o.$ stands for ``lower order terms''. The characteristic
speeds in the direction $n^i$ are given by $\beta^i n_i$, 
$\pm N + \beta^i n_i$, $\pm N\sqrt{\lambda_i} + \beta^i n_i$ (see next
subsection for a derivation and details),
where
\begin{eqnarray}
\lambda_1 &=& 2\sigma _{eff} \; ,
\nonumber\\
\lambda_2 &=& 1 + \chi - \frac{1}{2}(1 + \zeta)\eta + \gamma(2 - \eta + 2\chi) \; ,
\nonumber\\
\lambda_3 &=& \frac{1}{2}\,\chi + \frac{3}{8}(1 - \zeta)\eta
     -\frac{1}{4}(1 + 2\sigma _{eff})(\eta + 3\chi)\;  . \nonumber
\end{eqnarray}
The system is SH if
\begin{eqnarray*}
&& \lambda_j > 0, \;\;\; \hbox{for $j=1,2,3$}, \\
&& \lambda_3 = \frac{1}{4}(3\lambda_1 + 1) \;\;\;
  \hbox{if $\lambda_1 = \lambda_2$}. 
\end{eqnarray*}
The system has no superluminal speeds provided that $0 \leq \lambda _i\leq 1$. 
One way of achieving this is by asking all the $\lambda _i $'s to be $0$ or
$1$. Since $\lambda _i >0$ for strong hyperbolicity, one needs
$\lambda_i=1$. In particular, the condition $\lambda _1=1$ is 
possible only if the lapse is densitized with a constant $\sigma
_{eff}$ equal to $1/2$. In that case one has two families of formulations with speeds
along the light cone or normal to the hypersurfaces, one of them
bi-parametric and the other one mono-parametric, see \cite{kst}. If
$\sigma _{eff}$ is not constant, one can ask all the speeds but $\lambda _1
$ to be one. If one requires the parameters ${\zeta , \gamma , \eta ,
\chi }$ to be constant this leads to two mono-parametric families of SH systems:
\begin{displaymath}
\sigma _{eff}>0\, , \quad \zeta = -\frac{5}{3}, \gamma \mbox{
arbitrary }, 
\eta =\frac{6}{5} , \chi
= - \frac{2}{5} 
\end{displaymath}
or
\begin{displaymath}
\sigma _{eff}>0\, , \quad \zeta = -\frac{5 \chi + 8}{9\chi }, \gamma =
 -\frac{2}{3}, \eta =-3
\chi ,  \chi \ne 0 \mbox{ arbitrary }
\end{displaymath}
As shown below, the characteristic modes (and whether they are 
superluminal or not) associated with $\lambda _1$
are independent of the formulation, they depend only on the choice of
the slicing condition.

\subsection{Characteristic analysis}

Here, we discuss under which conditions the system (\ref{Eq:gij}-\ref{Eq:dkij})
is SH, and what the characteristic speeds are.
In order to do so, we choose a fixed direction $n^i$ with $g_{ij} n^i n^j = 1$
and study the eigenvalues of the 
principal part of (\ref{Eq:gij}-\ref{Eq:dkij})
in the direction of $n^k$:
\begin{equation}
\mu\left( \begin{array}{c} g_{ij} \\ K_{ij} \\ d_{kij} \end{array} \right)
 = \left( \begin{array}{ccc} 0 & 0 & 0 \\ 0 & 0 & \bbA \\ 0 & \bbB & 0 \end{array} \right)
   \left( \begin{array}{c} g_{ij} \\ K_{ij} \\ d_{kij} \end{array} \right),
\label{Eq-LinKd}
\end{equation}
where the matrices $\bbA$ and $\bbB$ are obtained from the principal parts of
(\ref{Eq:Kij}) and (\ref{Eq:dkij}), respectively, by replacing $\partial_k$ by $n_k$.
The characteristic speeds of the system are given by
\begin{displaymath}
N\mu + \beta^i n_i\; ,
\end{displaymath}
where $\mu$ are the eigenvalues of problem (\ref{Eq-LinKd}). The
system is SH if all the eigenvalues are real and the corresponding
eigenvectors are complete. 

These eigenvalues are either zero or can be obtained by considering
the eigenvalue problem
\begin{equation}
\mu^2 K_{ij} = \bbA\bbB K_{ij}\, .
\label{Eq-WaveK}
\end{equation}
Explicitly, we have
\begin{displaymath}
\mu^2 K_{ij} = K_{ij} + A n_{(i} n^s K_{j)s}
 + B n_i n_j K + C g_{ij} \left( n^r n^s K_{rs} - K\right),
\end{displaymath}
where the coefficients $A$, $B$ and $C$ are
\begin{eqnarray*}
A &=& -2 + \chi - \frac{3}{4}(\zeta - 1)\eta
 - \left(\sigma_{eff} + \frac{1}{2} \right)\left( \eta + 3\chi \right), \\
B &=& - \chi + \frac{3}{4}(\zeta - 1)\eta
   + \left(\sigma_{eff} + \frac{1}{2} \right)\left( 2 + \eta + 3\chi \right), \\
-2C &=& \chi - \frac{1}{2}(\zeta + 1)\eta + \gamma(2 - \eta + 2\chi).
\end{eqnarray*}

Next, we complete $n^i$ to a complex orthonormal basis $n^i$, $m^i$, $\bar{m}^i$
such that $n^i n^j g_{ij} = 1$, $m^i\bar{m}^j g_{ij} = 1$, and all other
scalar products are zero. We then decompose $K_{ij}$ according to
\begin{displaymath}
K_{ij} = a\, n_i n_j + b\, m_{(i}\bar{m}_{j)}
 + \left[ c\, n_{(i} m_{j)} + d\, m_{(i} m_{j)} + c.c. \right],
\end{displaymath}
where $a$, $b$ are real and $c$, $d$ are complex.
In this basis the linear operator $\bbA\bbB$
takes the simple form
\begin{equation}
\bbA\bbB = \left( \begin{array}{cccc}
1 + A + B & B - C  & 0               & 0 \\
        0 & 1 - 2C & 0               & 0 \\
        0 &      0 & 1 + \frac{A}{2} & 0 \\
        0 &      0 &               0 & 1
\end{array} \right).
\label{Eq-MainMat}
\end{equation}
The eigenvalues of this matrix are
\begin{eqnarray}
\lambda_1 &=& 1 + A + B = 2\sigma_{eff}, \nonumber\\
\lambda_2 &=& 1 - 2C = 1 + \chi - \frac{1}{2}(1 + \zeta)\eta + \gamma(2 - \eta + 2\chi),
\nonumber\\
\lambda_3 &=& 1 + \frac{A}{2} = \frac{1}{2}\,\chi + \frac{3}{8}(1 -
\zeta)\eta - \frac{1}{4}(1 + 2\sigma)(\eta + 3\chi),
\nonumber\\
\lambda_4 &=& 1, \nonumber 
\end{eqnarray}
and we have
\begin{displaymath}
B - C = \lambda_1 + \frac{1}{2}(\lambda_2 + 1) - 2\lambda_3\, .
\end{displaymath}

We demand that the matrix $\bbA\bbB$ be diagonalizable
and have only positive real eigenvalues. As we show now, this is  
a sufficient condition for
the original system (\ref{Eq:gij}-\ref{Eq:dkij}) to be SH. 

First note that since $\bbA\bbB$ is positive definite, 
$\bbB$ is injective and $\bbA$ is surjective. Now let $K^{(1)}_{ij}\,
, ... , K^{(6)}_{ij}$ denote the six real eigenvectors of
$\bbA\bbB$ which have the eigenvalues $\omega^{(1)}, ..., \omega^{(6)}$.
Then, the $12$ vectors
\begin{equation}
v_{\pm s} = \left( \begin{array}{c} 0 \\ \pm\sqrt{\omega^{(s)}} K^{(s)}_{ij} \\ (\bbB K^{(s)})_{kij}
\end{array} \right),
\label{Eq-EV}
\end{equation}
are eigenvectors of (\ref{Eq-LinKd}) with eigenvalues
$\mu_{\pm s} = \pm\sqrt{\omega^{(s)}}$, $s = 1,2,...,6$.
These vectors are linearly independent since $\omega^{(s)}\neq 0$
and since $\bbB$ is injective. The remaining eigenvectors have $\mu=0$.
Since $\bbA$ is surjective, we have
\begin{displaymath}
\dim\ker\bbA = 18 - \dim\Im\bbA = 12,
\end{displaymath}
thus there are $18$ zero eigenvectors, $12$ with non-trivial $d_{kij}$'s
which lie in the kernel of $\bbA$, and $6$ with non-trivial $g_{ij}$'s.
Therefore, we have a set of $30$ independent eigenvectors with real
eigenvalues. In order to show  that the system
(\ref{Eq:gij}-\ref{Eq:dkij}) yields a well posed initial-value
problem, one has to look at the matrix
$S(n^i)$ whose columns are the $30$ eigenvectors of the principal part
of the system and show that $S(n^i)$ and its inverse are uniformly
bounded and that they depend smoothly on
$n^i$ and the metric coefficients $g_{ij}$ and $g^{ij}$. We do not show
this.
However, if we linearize the equations around flat spacetime in Cartesian coordinates,
the principal part depends only on $n^i$ and the flat metric $\delta_{ij}$.
Using an isotropy argument it is not difficult to show that in this case
the matrix $S(n^i)$ can be obtained from $S(n_0^i)$ in a fixed direction
$n_0^i$ by a rotation which maps $n_0^i$ to $n^i$. It follows from this
that $S(n^i)$ can be chosen such that its norm and the norm of its
inverse are uniformly bounded for all $n^i$ with $\delta_{ij} n^i n^j = 1$.
In this case, this is sufficient to guarantee well posedness
\cite{Kreiss-Lorenz}.

The matrix $\bbA\bbB$ is diagonalizable if and only if
$\lambda_1 , \lambda_2 , \lambda_3 > 0$ and
$\lambda_3 = (3\lambda_1 + 1)/4$ whenever $\lambda_1 = \lambda_2$.

\subsection{Characterization of the eigenmodes}
Here we give an interpretation to the
eigenmodes of (\ref{Eq-WaveK}) when the SH field equations
(\ref{gdot},\ref{kdot},\ref{ddot}) 
are linearized around Minkowski spacetime. One can see that in such case the
characteristic modes are
solutions to Einstein's equations. According to (\ref{Eq-MainMat})
these modes are given by the following vectors corresponding to the values
of $(a,b,c,d)$:
\begin{eqnarray}
\lambda = 1: && (0,0,0,1),(0,0,0,i), \nonumber\\
\lambda = \lambda_3: && (0,0,1,0),(0,0,i,0), \nonumber\\
\lambda = \lambda_2: && (0,1,0,0) \mbox{ if } \lambda_2=
\lambda_1  ,\quad 
 (C-B, 2\sigma _{eff}-1+2C, 0,0) \mbox { otherwise } ,
\quad\nonumber\\
\lambda = \lambda_1: && (1,0,0,0). \label{gauge_mode}
\end{eqnarray}
The first two modes, which propagate along the light cone (ie. which
have speeds $\pm N + \beta^i n_i$) are physical modes: 
By making a Fourier transform of the linearized Hamiltonian and
momentum constraints, it is easy to check that the constraints are satisfied by
the corresponding 4 eigenmodes
(\ref{Eq-EV}). 
Next, we have six constraint violating modes which have
characteristic speeds $\pm N\sqrt{\lambda_3} + \beta^i n_i$,
$\pm N \sqrt{\lambda_2} + \beta^i n_i$:
Indeed, the Fourier transformed linearized momentum constraint yields
\begin{displaymath}
n^j K_{ij} - n_i K = -b\, n_i + \frac{1}{2}\, c\, m_i + \frac{1}{2}\,\bar{c}\,\bar{m}_i\, ,
\end{displaymath}
and only modes with $b = c = 0$ satisfy these constraints.
The modes with characteristic speeds $\pm N\sqrt{2\sigma _{eff}} +
\beta^i n_i$ are  
gauge modes: With respect to an infinitesimal coordinate transformation
of the form $\delta x^\mu \mapsto \delta x^\mu + f\delta^\mu_t\,$, the
extrinsic curvature transforms according to
\begin{displaymath}
K_{ij} \mapsto K_{ij} + n_i n_j f\,  ,
\end{displaymath}
therefore 
\begin{displaymath}
a \mapsto a + f \, ,
\end{displaymath}
and the eigenmode (\ref{gauge_mode}) with eigenvalue $\lambda _1 = 
2\sigma _{eff}$ can be gauged away.

\section{Strongly hyperbolic formulations with a live gauge}
\label{live_section}

Here we construct SH formulations with live gauges. 
In addition to the variables of the previous section,
we promote the lapse $N$ and three extra quantities $A_i$
which are equal to $(\partial_i N)/N$ if the constraints are
satisfied, to independent variables. 
Our variables are, therefore, $\{g_{ij}, K_{ij}, d_{kij}, N, A_i\}$. 
As in the previous section, the shift is assumed to be an arbitrary
but apriori prescribed function of spacetime.

As evolution equation for the lapse we consider
\begin{displaymath}
\partial_0 N = - F(N,K,x^{\mu} ),
\end{displaymath}
where $F(N,K,x^{\mu} )$ is an arbitrary function of its
arguments. An evolution equation for $A_i$ is obtained from
this by taking a spatial derivative:
\begin{displaymath}
\partial_0 A_i = -\frac{\partial F}{\partial N} A_i 
- \frac{1}{N}\frac{\partial F}{\partial K} \partial_i K 
- \frac{1}{N}\frac{\partial F}{\partial x^i} + \xi \,  C_i\, ,
\end{displaymath}
where we have also added the momentum constraint with a free parameter $\xi$.

The evolution equations for $g_{ij}$, $K_{ij}$ and $d_{kij}$ are the same
as in (\ref{gdot}-\ref{ddot}) where now
\begin{displaymath}
\frac{1}{N} \nabla_i\nabla_j N = \partial_{(i} A_{j)} - \Gamma^k_{\;
ij} A_k + A_i A_j\; ,
\end{displaymath}
and where we replace $(\partial k N)/N$ by $A_k$ in the evolution
equation for $d_{kij}$.

The principal part of the system is
\begin{eqnarray*}
\partial_0 g_{ij}&=&  l.o.\, , \\
\partial _0 K_{ij}&=& \frac{1}{2} g^{ab}\left( 
  - \partial_a d_{bij} + (1+\zeta )\partial_a d_{(ij)b} 
  + (1-\zeta)\partial_{(i} d_{|ab|j)} - \partial_{(i}d_{j)ab} 
  + \gamma\, g_{ij} g^{kl}\partial_a(d_{klb}-d_{bkl}) \right) 
  - \partial_{(i}A_{j)} + l.o. \, , \\
\partial _0 d_{kij}&=& -2\partial _kK_{ij} 
 + \eta\,g_{k(i}g^{ab}(\partial_{|a|} K_{j)b} - \partial_{j)} K_{ab}) 
 + \chi\,g_{ij}g^{ab}(\partial_a K_{kb} - \partial_k K_{ab}) + l.o. \, , \\
\partial_0 N &=& l.o. \, , \\
\partial_0 A_i &=& -\frac{1}{N}\frac{\partial F}{\partial K}\partial _iK
+ \xi\, g^{ab}(\partial_a K_{bi}-\partial_i K_{ab}) + l.o. \, .
\end{eqnarray*}

In order to get the conditions under which the system is SH,
and in order to get the characteristic speeds, we can use the techniques
of the previous section: The principal part has the form
\begin{displaymath}
\mu\left( \begin{array}{c} N \\ g_{ij} \\ K_{ij} \\ u \end{array} \right)
 = \left( \begin{array}{cccc} 0 & 0 & 0 & 0 \\ 0 & 0 & 0 & 0 \\ 0 & 0 & 0 & \bbA \\ 0 & 0 & \bbB & 0 \end{array} \right)
   \left( \begin{array}{c} N \\ g_{ij} \\ K_{ij} \\ u \end{array} \right),
\end{displaymath}
where $u = (d_{kij}, A_i)^T$.
Explicitly writing the matrix $\bbA\bbB$, we obtain the same result as
in (\ref{Eq-MainMat}) except that now
\begin{eqnarray*}
A &=& -2 - \frac{1}{2}\chi - \frac{1}{4}(3\zeta -1)\eta - \xi \\
B &=& \frac{1}{2}\chi + \frac{1}{4}(3\zeta -1)\eta + \xi + 1 +
2\sigma _{eff} \\
-2C &=& \chi - \frac{1}{2}(\zeta + 1)\eta + \gamma(2 - \eta + 2\chi)\, , 
\end{eqnarray*}
where $\sigma_{eff}  = (\partial_KF)/(2N)$. Therefore, the characteristic 
speeds of the system are 
$\beta^i n_i$, $\pm N + \beta^i n_i$, $\pm N\sqrt{\lambda_i} + \beta^i n_i$, 
with
\begin{eqnarray*}
\lambda _1 &=& 2\sigma_{eff}\, ,\\
\lambda _2 &=& 1 + \chi - \frac{1}{2}(1 + \zeta)\eta + \gamma (2-\eta + 2\chi),\\
\lambda _3 &=& -\frac{1}{4}\chi - \frac{1}{8}(3\zeta - 1)\eta - \frac{1}{2}\xi\,.
\end{eqnarray*}
In particular, if one chooses
$\xi = \sigma_{eff} (\eta + 3\chi )$, these speeds are exactly those of the
system of the previous section and strong hyperbolicity holds under
the same conditions. The conditions under which there are no
superluminal speeds are also the same, and the mode associated with
$\lambda _1$ is related to the choice of gauge as well. More
generally, one can, as in the previous section, 
ask all the $\lambda _i$'s but $\lambda _1$ to be $1$.
This leads to two SH many-parameter families, one with three free
parameters ($ \gamma , \zeta , \eta $):
\begin{eqnarray*}
\sigma _{eff} &>&0 \,, \\
\gamma & \neq & -\frac{1}{2}\,, \\
\chi & = & \frac{(1+\zeta )\eta - 2 \gamma (2-\eta)}{2(1+2\gamma )} \,,\\
\xi & = & -\frac{1}{2}\chi - \frac{1}{4}(3\zeta -1)\eta - 2\,,
\end{eqnarray*}
and another one with two free parameters ($\zeta , \chi $):
\begin{displaymath}
\sigma _{eff} >0 \,, \quad
\gamma = -\frac{1}{2}\, , \quad
\zeta \eta = -2 \, , \quad
\xi= -\frac{1}{2}\chi + \frac{1}{4}\eta - \frac{1}{2} \; .
\end{displaymath}

\section{Some symmetric hyperbolic subfamilies} \label{symm}

Here we show that some of our formulations are  symmetric
hyperbolic. We do not intend to give the most general conditions under
which this holds but, instead, show
that there are at least some subfamilies that are symmetric hyperbolic. 
We start discussing the family of live
gauge conditions and then briefly discuss the algebraic case. 

We show that the principal part of the system
can be brought into symmetric form by using a transformation which
does not depend on $n^i$. In order to find such a transformation, it
is convenient to first transform the variables $K_{ij}$ and
$d_{kij}$ into their trace and trace-less parts:
\begin{eqnarray}
K_{ij} &=& P_{ij} + \frac{1}{3}\, g_{ij} K, \nonumber\\
d_{kij} &=& 2e_{kij} + \frac{3}{5}\, g_{k(i}\Gamma_{j)} - \frac{1}{5}\, g_{ij}\Gamma_k + \frac{1}{3}\, g_{ij} d_k\, ,
\nonumber
\end{eqnarray}
where $\Gamma_k = b_k - d_k/3$ and $P_{ij}$ and $e_{kij}$ are trace-less
in all their indices. In terms of the new variables $K$, $A_{ij}$,
$e_{kij}$, $\Gamma_k$, $d_k$, the principal part is
\begin{eqnarray}
\mu K &=& \left( 1 + \frac{3}{2}\gamma \right) \left( \Gamma_n - \frac{2}{3} d_n \right) - A_n\; ,
\nonumber\\
\mu P_{ij} &=& -e_{nij} + (1+\zeta)e_{(ij)n} 
 + \left[ \frac{1}{20}(5-9\zeta) n_{(i} \Gamma_{j)} - \frac{1}{6}\, n_{(i} d_{j)} - n_{(i} A_{j)} \right]^{TF},
\nonumber\\
\mu \Gamma_k &=& \left( \frac{5}{3}\,\eta - 2 \right) P_{kn} - \frac{10}{9}\,\eta\, n_k K,
\nonumber\\
\mu d_k &=& (\eta + 3\chi) P_{kn} - \frac{2}{3}(3 + \eta + 3\chi) n_k K,
\nonumber\\
\mu A_k &=& \xi P_{kn} -\left(2\sigma_{eff} + \frac{2}{3}\xi \right) n_k K,
\nonumber\\
\mu e_{kij} &=& -n_k P_{ij} + \frac{3}{5}\, g_{k(i} P_{j)n} - \frac{1}{5}\, g_{ij} P_{kn}\, ,
\nonumber
\end{eqnarray}
where $TF$ indicates the trace-free part, and where $\Gamma_n = n^k \Gamma_k$,
$e_{nij} = n^k e_{kij}$ etc. 
First, we see that if we set $\zeta = -1$, the term involving $e_{kij}$ in the 
equation for $P_{ij}$ is the symmetric counterpart of the term involving $P_{ij}$ 
in the equation for $e_{kij}$. Now, we can try to find three independent linear combinations
of the variables $\Gamma_k$, $d_k$ and $A_k$ and rescale $K$ such that
the principal part becomes symmetric.
For the choice $\zeta = -1$, $\gamma = -2/3$, $\eta = 6m/5$, $\chi = -2m/5$, we
find that the transformation
\begin{displaymath}
K \mapsto c_1 K, \qquad
\Gamma_k \mapsto c_2\left(\Gamma_k - \frac{2m}{3} d_k \right), \qquad
d_k \mapsto c_3\left[ \left(\sigma_{eff} + \frac{1}{3}\xi\right) d_k -
A_k \right], \qquad
A_k \mapsto c_4 A_k\, ,
\end{displaymath}
yields a symmetric system, provided that
\begin{displaymath}
m > 1, \qquad
\sigma_{eff} >  p \equiv \frac{7m}{15} - \frac{1}{6}\, ,
\end{displaymath}
and if we choose
\begin{eqnarray}
\xi &=& \frac{1}{2c_4^2} \left[ -(1+3c_4^2\sigma_{eff}) + \sqrt{ (1-3c_4^2\sigma_{eff})^2 + 12c_4^2p} \right],
\nonumber\\
c_1^2 &=& 2c_4^2\left(\sigma_{eff} + \frac{1}{3}\xi\right), \nonumber\\
c_2^2 &=& \frac{7}{20(m-1)}\, ,\nonumber\\
c_3^2 &=& -\frac{1 + c_4^2\xi}{\xi}\, . \nonumber
\end{eqnarray}
Note that the definition of $\xi$ and the requirement  $\sigma_{eff} >
p$ guarantee that  $-1 < c_4^2\xi < 0$ and $3\sigma _{eff}+\xi >0$.
The characteristic speeds of the system are
$\beta^i n_i$, $\pm N + \beta^i n_i$, $\pm N\sqrt{\lambda_i} + \beta^i n_i$,
with
\begin{displaymath}
\lambda_1 = 2\sigma_{eff}, \qquad
\lambda_2 = 2p, \qquad
\lambda_3 = \frac{7m}{10} - \frac{1}{2}\xi .
\end{displaymath}
Setting, for example, $m = 15/14$ and $c_4^2=2$ yields $\lambda_2 = 2/3$ and 
$\lambda_3 = 3/4 - \xi/2 < 1$. Therefore, as long as $\sigma_{eff} > 1/3$,
the system is symmetric hyperbolic and has no superluminal constraint
violating modes.

One can similarly show that there are symmetric hyperbolic
subfamilies in the algebraic case. 
For example, the choice
\begin{displaymath}
\zeta = -1, \qquad
\gamma = -\frac{2}{3}\, \qquad
\eta = \frac{3}{7}(1 + 6\sigma_{eff}), \qquad
\chi = -\frac{1}{7}(1 + 6\sigma_{eff}),
\end{displaymath}
yields a symmetrizable system provided that
$\sigma_{eff} > 3/10$. In this case the $\lambda_i$'s are
$\lambda_1 = \lambda_2 = 2\sigma _{eff}$, $\lambda_3 = (1 + 6\sigma_{eff})/4$.

Finally, as in KST, one could perform a many-parameter change of variables
in any of our formulations
without affecting the spectrum of the principal part and therefore
without changing the level of hyperbolicity.

\section{Discussion} \label{discussion}

The freedom in the KST formulations has proven to be very useful in
improving the stability of 3D single black hole numerical
evolutions. There are studies
underway to understand the reasons behind this, but there is still not
a clear picture. If the lifetime of present evolutions using this
system carry over to dynamical situations, interesting parts of a binary black hole
collision could be described. However, one possible obstacle is the
lack of flexibility of the family of lapses considered in the KST
formulation, namely the densitized lapse prescription.  On the other
hand, the main idea behind the BM formulation is to
introduce in SH formulations several of the dynamical
slicings that are used in numerical evolutions. The
purpose of this paper has been to combine the spirits of the KST and
BM formulations, and the result is expected to be useful in 3D
evolutions, especially in dynamical ones. 

We have considered an algebraic and a live
family of slicing conditions that include most
of the conditions used in numerical relativity, such as densitized
lapse, time harmonic slicings, the ``$1+\log$'' case,
and the BM conditions. Furthermore, we have shown that one can obtain
symmetric hyperbolic systems with these choices of gauges and,
therefore, have a well posed initial-value problem. Up to our
knowledge, this is the first time this has been achieved. The BM
formulation, for example, is shown in \cite{bona} to be SH in the
sense used in this paper but, as already mentioned, this does not
automatically imply well posedness. 

Initially the approach to black hole evolutions (see
\cite{leh} for a review in numerical relativity) was through the use of
singularity avoiding (SA) slicings, such as maximal slicing. This gauge
condition has been widely used in 1D
and 2D, but in 3D several difficulties appear: not only is it
computationally expensive to solve elliptic equations at each time
step, but also  one has to solve this elliptic 
equation with very good accuracy in order to
avoid noise (see, e.g. \cite{ann}). 
This lead 
to the introduction of
live gauges that mimic the maximal condition near the singularity. All of the
BM conditions are SA exactly in those cases that lead to SH
formulations \cite{bona1}, but there are still differences between different
subcases. For example, time harmonic slicings are not as SA \cite{bona2} as the
``$1+\log$'' are, and this is one of the reasons the latter has been used
so much. However, if the singularity is not avoided but is excised, SA
slicings are in principle no longer needed. On the contrary, these
slicings can cause problems since they 
introduce steep gradients near the horizon. Still, there is evidence that the
use of an appropriate shift can help,  making the 
``$1+\log$'' condition useful even in the presence of excision \cite{alc}.

Having a hyperbolic formulation with no
superluminal speeds ($\lambda_i \leq 1$) is an advantage for singularity excision since
then one does not need to give boundary conditions at the inner boundary.
In the SH families of formulations we have presented in this paper, all the $\lambda_i$'s
but one ($\lambda_1$) can apriori be set to $1$. As we have shown, 
the modes associated with $\lambda_1$ are gauge modes, and thus,
whether these modes are superluminal or not does not depend on the
formulation but on the gauge condition considered.
In some cases one can decide a priori whether or not $\lambda_1\leq
1$, but in principle $\lambda_1$ might depend on the solution.
If that happens, one can ask the code to follow the characteristic speeds
and decide during evolution whether boundary conditions at the inner
boundary are needed or not (at the outer boundary one has to deal with
boundary conditions in any case). If a mode does enter the computational
domain and one ignores this fact and continues, for instance, by doing 
extrapolation,
 one is implicitly giving boundary conditions that
depend on the grid spacing and might not have a consistent limit or be
physically correct as
resolution is increased. This can indeed happen even with very simple
choices of gauge (see, e.g. the appendix of \cite{cal1}) and the result
of such a procedure is uncertain. Even if one
is willing to give boundary conditions to modes that enter the
domain, the question of what conditions to give still remains. In principle,
since physically relevant quantities are gauge-independent, one could
give any boundary conditions to these modes. However, what could
happen is that a bad choice leads to a gauge the becomes singular
after a while. In fact, the same problem arises even in
the absence of boundaries when the lapse is not prescribed apriori
as a function of spacetime, but, for instance, a live gauge condition
is chosen \cite{gauge}.

An issue that we have not addressed in this paper is the introduction
of dynamical shifts that may help to follow steep gradients near the horizon
or provide some sort of minimal distortion. Current efforts are
oriented along this line.  Numerical experiments testing the
formulations presented in this paper are also underway.

\section{Acknowledgments}
This work was supported in part by NSF grant PHY9800973, the Horace C. Hearne
Jr. Institute of Theoretical Physics, the Swiss National Science
Foundation, and Fundaci\'on Antorchas. We thank Carles Bona, Luis Lehner, Jorge
Pullin, and Mark Scheel for useful comments and discussions. 
During the preparation of this manuscript related work by Alvi
has appeared \cite{alvi}.



\end{document}